\documentclass[12pt,a4paper]{article}
\usepackage{epsfig}
\begin{document}
\title{Neutral top-pion $ \pi_{t}^{0}$ and  $t\gamma(z)$ production at the HERA and THERA colliders}
\author{Chongxing Yue, Dongqi Yu, Zhengjun Zong\\
{\small Department of Physics, Liaoning  Normal University, Dalian
116029. P.R.China}
\thanks{E-mail:cxyue@lnnu.edu.cn}}
\date{\today}
\maketitle
\begin{abstract}
\hspace{5mm}In the context of top-color assisted technicolor
($TC2$) models, we calculate the contributions of the neutral
top-pion $ \pi_{t}^{0}$ to $t\gamma$  and $tz$ production via the
processes $ ep\rightarrow \gamma c \rightarrow t\gamma $ and $
ep\rightarrow \gamma c \rightarrow tz$ at the $HERA$ and $THERA$
colliders. Our results show that the cross sections
 $\sigma_{t\gamma}(s)$ and $\sigma_{tz}(s)$ are very small at
the $HERA$  collider with $\sqrt{s}=320GeV $. However, in most of
the parameter space, $\sigma_{t\gamma}(s)$ or $\sigma_{tz}(s)$ is
in the range of about $0.1pb \sim 1 pb$ at the $THERA$ collider
with $\sqrt{s}=1000GeV$.
\end {abstract}
\newpage
Single top quark production is very sensitive to the anomalous top
quark couplings $tqv$, in which $q$ represents the up quark or
charm quark and $v$ represents the gauge bosons $z$, $\gamma$, or
$g$[1]. This type of couplings can be generated in supersymmetry,
 top-color scenario, and other specific models beyond the
standard model($SM$). Thus, studying the contributions of the
$tqv$ couplings to single top production is of interest. This fact
has led to many studies involving single top production via the
$tqv$ couplings in lepton colliders[2,3] and hadron
colliders[4,5].

The $HERA$ collider and the $THERA$ collider[6] with the
center-of-mass energy $\sqrt{s}=320GeV$ and $1000 GeV$,
respectively, are the experimental facilities where high energy
electron-proton and positron-proton interactions can be studied.
It is well known that, in the $SM$, single top quark can not be
produced at an observable rate in these high energy colliders.
However, it has been shown that the $HERA$ collider and the
$THERA$ collider can provide a very good sensitivity on the $tqv$
couplings via single top production[7]. This type of single top
production may be detected in these colliders[5,8]. The $HERA$ and
$THERA$ colliders are powerful tools for searching for the
anomalous top quark couplings $tqv$.

The presence of the top-pions $\pi_{t}^{0,\pm}$ in low-energy
spectrum is an inevitable feature of top-color scenario[9]. These
new particles have large Yukawa couplings to the third family
quarks and can induce the tree-level flavor changing($FC$)
couplings, which have significant contributions to the anomalous
top quark couplings $tqv$[10]. In Ref.[5] we study the
contributions of the $tqv$ couplings generated by $\pi^{0}_{t}$
exchange to single top production via the t-channel process
$eq\rightarrow$ et at the $HERA$ and $THERA$ colliders. We have
shown that it can generate significant effects on the process $ec
\rightarrow$ et, which may be observable in the $THERA$ collider.
The aim of this Letter is to consider the contributions of the
$tcv$ couplings given by $\pi^{0}_{t}$ exchange to the processes
$ep\rightarrow \gamma c \rightarrow t \gamma$ and $ep\rightarrow
\gamma c\rightarrow tz$ in the context of topcolor-assisted
technicolor ($TC2$) models[11], and see whether the effects of the
neutral top-pion $\pi^{0}_{t}$ on $t\gamma$ and $tz$ production
can be detected in the $HERA$ collider or the $THERA$ collider.

For $TC2$ models[9,11], the underlying interactions, topcolor
interactions, are assumed to be chiral critically strong at the
scale about 1 $TeV$ and coupled preferentially to the third
generation, and therefore do not posses $GIM$ mechanism. The
non-universal gauge interactions result in the tree-level $FC$
coupling vertices when one writes the interactions in the mass
eigen-basis, which can induce the anomalous top quark couplings
$tuv$ and $tcv$. However, the $tuv$ couplings can be neglected
because the $FC$ scalar coupling $\pi^{0}_{t}$ $tu$ is very small
[12]. The effective forms of the $tc \gamma$ and $tcz$ coupling
vertices $\Lambda_{tc\gamma}$, $\Lambda_{tcz}$ have been given in
$Eq$.[4] and $Eq$.[5] of Ref.[10].

\begin{figure}[htb]
\vspace{-2.5cm}
\begin{center}
\epsfig{file=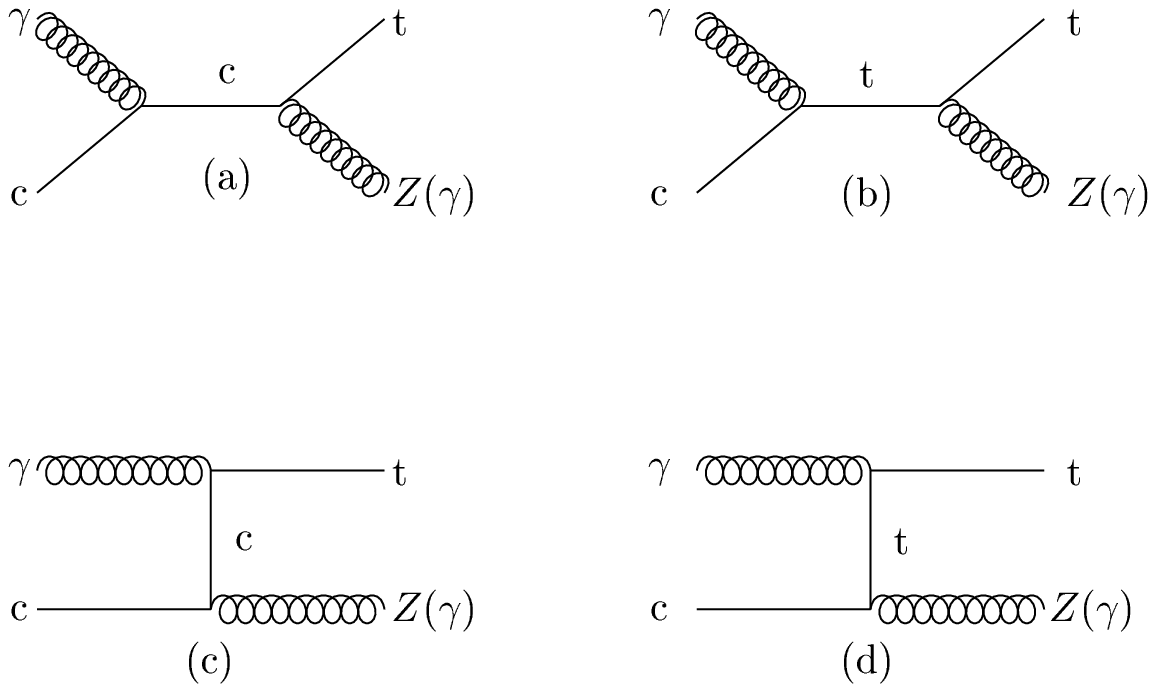,width=550pt,height=800pt} \vspace{-18.5cm}
\hspace{1cm} \caption{Feynman diagrams for $t\gamma(z)$ production
contributed by the anomalous top \hspace*{2.0cm}coupling vertices
$\Lambda_{tc\gamma}(\Lambda_{tcz})$.} \label{ee}
\end{center}
\end{figure}

From the above discussion, we can see that the neutral top-pion
$\pi^{0}_{t}$ can generate contributions to the subprocesses
$\gamma c\rightarrow t\gamma$ and $\gamma c\rightarrow tz$ via the
anomalous top quark couplings $tc\gamma$ and $tcz$ generated by
the $\pi^{0}_{t}$ $\bar{t}c$ coupling. The relevant Feynman
diagrams are shown in Fig.1.

For the subprocesses $c(p_{c})+\gamma(p_{\gamma}) \rightarrow
t(p_{t})+\gamma(p_{\gamma}')$ and
$c(p_{c})+\gamma(p_{\gamma})\rightarrow t(p_{t}')+z(p_{z})$, we
define the kinematical invariants
$\hat{s}=(p_{t}+p_{\gamma}')^{2}=(p_{t}'+p_{z})^{2},
t=(p_{\gamma}-p_{t})^{2},$ and $t'=(p_{\gamma}-p_{t}')^{2}$. The
renormalized amplitudes for these processes can be written as:
\begin{eqnarray}
M^{t\gamma}=M_{s}^{t\gamma}+M_{t}^{t\gamma}&=&
\bar{u}_{t}\Lambda_{tc\gamma}^{\mu}i\varepsilon_{\mu}
\frac{i[\gamma\cdot(p_{t}+p_{\gamma}')+m_{c}]}{\hat{s}-m_{c}^{2}+i\mu}i
\varepsilon^{\nu}(\frac{2}{3}ie\gamma_{\nu})u_{c}\nonumber\\
&&+\bar{u}_{t}(\frac{2}{3}ie\gamma^{\mu})i\varepsilon_{\mu}\frac{i[\gamma
\cdot(p_{t}+p_{\gamma}')+m_{t}]}{\hat{s}-m_{t}^{2}+im_{t}\Gamma_{t}}i
\varepsilon^{\nu}\Lambda_{tc\gamma,\nu}u_{c}\nonumber\\
&&+\bar{u}_{t}\Lambda_{tc\gamma}^{\mu}i\varepsilon_{\mu}\frac{i[\gamma\cdot(p_{\gamma}-p_{t})
+m_{c}]}{t-m_{c}^{2}+i\mu}i
\varepsilon^{\nu}(\frac{2}{3}ie\gamma_{\nu})u_{c}\nonumber\\
&&+\bar{u}_{t}(\frac{2}{3}ie\gamma^{\mu})i\varepsilon_{\mu}\frac{i[\gamma
\cdot(p_{\gamma}-p_{t})+m_{t}]}{t-m_{t}^{2}+im_{t}\Gamma_{t}}i
\varepsilon^{\nu}\Lambda_{tc\gamma,\nu}u_{c},
\end{eqnarray}
\begin{eqnarray}
M^{tz}=M_{s}^{tz}+M_{t}^{tz}&=&
\bar{u}_{t}\Lambda_{tc\gamma}^{\mu}i\varepsilon_{\mu}
\frac{i[\gamma\cdot(p_{t}'+p_{z})+m_{c}]}{\hat{s}-m_{c}^{2}+i\mu}i
\varepsilon^{\nu}(\frac{2}{3}ie\gamma_{\nu})u_{c}\nonumber\\
&&+\bar{u}_{t}\Lambda_{tcz}^{\mu}i\varepsilon_{\mu}\frac{i[\gamma
\cdot(p_{t}'+p_{z})+m_{t}]}{\hat{s}-m_{t}^{2}+im_{t}\Gamma_{t}}i
\varepsilon^{\nu}\Lambda_{tc\gamma,\nu}u_{c}\nonumber\\
&&+\bar{u}_{t}\Lambda_{tc\gamma}^{\mu}i\varepsilon_{\mu}\frac{i[\gamma\cdot(p_{\gamma}-p_{t}')
+m_{c}]}{t'-m_{c}^{2}+i\mu}i
\varepsilon^{\nu}\Lambda_{zc\bar{c},\nu}u_{c}\nonumber\\
&&+\bar{u}_{t}(\frac{2}{3}ie\gamma^{\mu})i\varepsilon_{\mu}\frac{i[\gamma
\cdot(p_{\gamma}-p_{t}')+m_{t}]}{t'-m_{t}^{2}+im_{t}\Gamma_{t}}i
\varepsilon^{\nu}\Lambda_{tcz,\nu}u_{c}
\end{eqnarray}
with
$$\Lambda_{zc\bar{c}}^{\mu}=\Lambda_{zt\bar{t}}^{\mu}=\frac{e}{s_{W}c_{W}}
[(\frac{1}{2}-\frac{2}{3}s_{W}^{2})\gamma^{\mu}\frac{1-\gamma_{5}}{2}-
\frac{2}{3}s_{W}^{2}\gamma^{\mu}\frac{1+\gamma_{5}}{2}].$$ Where
$\mu$ is a real parameter, which is introduced to make the
integral convergent. $\Gamma_{t}$ is the total decay width of the
top quark.

After calculating the cross section $\hat{\sigma}_{i}(\hat{s})$ of
the subprocesses $\gamma c\rightarrow t\gamma$ or $\gamma
c\rightarrow tz$, the total cross section $\sigma_{i}(s)$ of
$t\gamma$ production or $tz$ production can be obtained by folding
$\hat{\sigma}_{i}(\hat{s})$ with the charm quark distribution
$f_{c/p}(x)$ in proton and the backscattered high energy photon
spectrum $f_{\gamma/e}(\frac{\tau}{x})$:$$\sigma_{i}(s)=
\int^{0.83}_{\tau_{min}}d\tau\int^{1}_{\tau/0.83}\frac{dx}{x}
f_{\gamma/e}(\frac{\tau}{x})f_{c/p}(x)\hat{\sigma}_{i}(\hat{s})$$
with $\hat{s}=\tau s,\ \ \tau_{min}=\frac{m_{t}^{2}+m_{v}^{2}}{s}$
and $f_{\gamma/e}(x)$ can be written as [13]:
$$f_{\gamma/e}(x)=\frac{1}{1.84}[1-x+\frac{1}{1-x}[1-\frac{4x}{x_{0}}
(1-\frac{x}{x_{0}(1-x)})]] \ \ \ (x_{0}=4.83).$$ The parton
distribution function $f_{c/p}(x)$ of the charm quark runs with
the energy scale. In our calculation, we will take the CTEQ5
parton distribution function for $f_{c/p}(x)$[14].

\begin{figure}[htb]
\vspace{-0.5cm}
\begin{center}
\epsfig{file=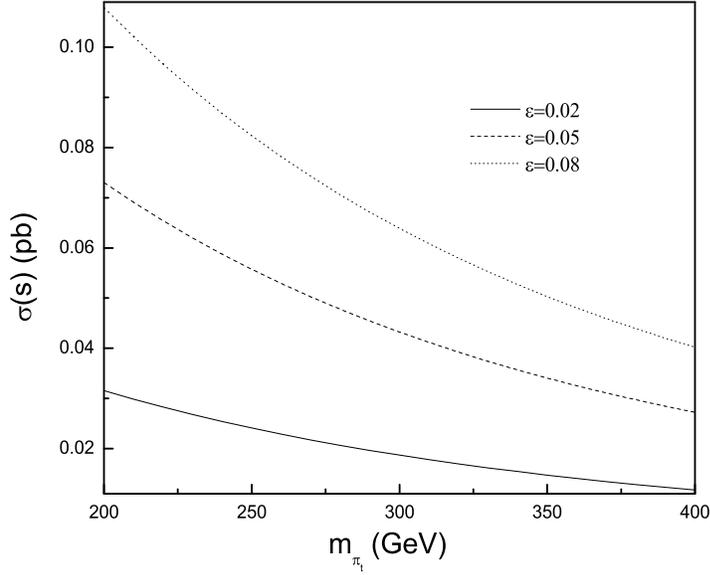,width=300pt,height=250pt} \vspace{-1.0cm}
\hspace{5mm} \caption{The cross section $\sigma_{t\gamma}(s)$ as a
function of the top-pion mass $m_{\pi_{t}}$ for three values
\hspace*{1.8cm}of the parameter $\varepsilon$ at the HERA
collider.} \label{ee}
\end{center}
\end{figure}
\vspace*{11cm}

\begin{figure}[htb]
\vspace*{-12.5cm}
\begin{center}
\epsfig{file=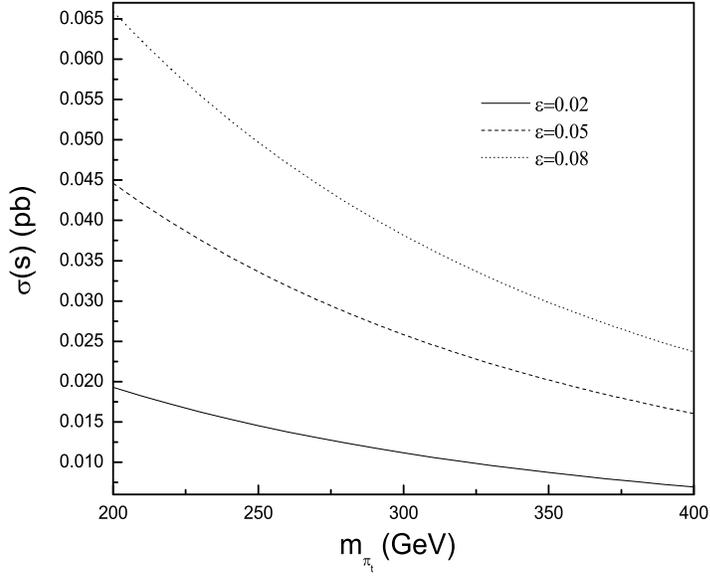,width=300pt,height=250pt} \vspace{-1.0cm}
\hspace{5mm} \caption{Same as Fig.2 but for the cross section
$\sigma_{tz}(s)$}\label{ee}
\end{center}
\end{figure}

To obtain numerical results, we take the fine structure constant
$\alpha_{e}=\frac{1}{128.8},$ $m_t=175GeV$, $m_c=1.2GeV$,
$m_z=91.2GeV$, and assume that the total decay width $\Gamma_{t} $
of the top quark is dominated by the decay channel $t\rightarrow
wb$, which has been taken  $\Gamma (t\rightarrow wb)=1.56GeV$
[15]. The limits on the mass $m_{\pi_{t}}$ of the top-pion
$\pi^{0}_{t}$ may be obtained via studying it's effects on
observables[9]. It has been shown that $m_{\pi_{t}}$ is allowed to
be in the range of a few hundred $GeV$ depending on the models.
For $TC2$ models, top-color interactions make small contributions
to electroweak symmetry breaking and give rise to the main part of
the top quark mass, (1-$\varepsilon$)$m_{t}$, with the parameter
$\varepsilon\ll1$. As numerical estimation, we will take
$m_{\pi_{t}}$ and $\varepsilon$ as free parameters.

\begin{figure}[htb]
\vspace{-0.5cm}
\begin{center}
\epsfig{file=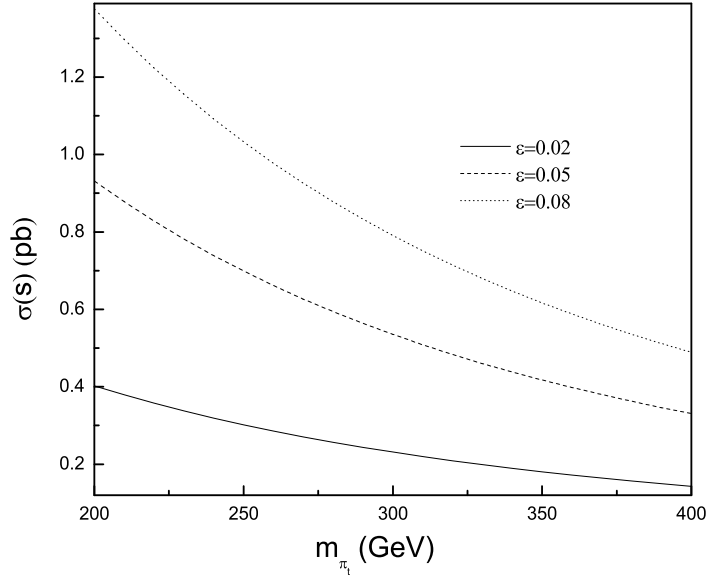,width=300pt,height=250pt} \vspace{-1.0cm}
\hspace{5mm} \caption{The cross section $\sigma_{t\gamma}(s)$ as a
function of $m_{\pi_{t}}$ for three values of the parameter
\hspace*{1.8cm} $\varepsilon$ at the THERA collider.} \label{ee}
\end{center}
\end{figure}

Our numerical results are summarized in Fig.2-Fig.5. From these
figures, we can see that the cross sections $\sigma_{t\gamma}(s) $
and $ \sigma_{tz}(s)$ of $t\gamma$ and $tz$ production at the
$HERA$ and $THERA$ colliders increase as the parameter
$\varepsilon$ increases and  $m_{\pi_{t}}$ decreases. In all of
the parameter space, we have that the cross section
$\sigma_{t\gamma}(s)$ of the process $ep\rightarrow\gamma
c\rightarrow t\gamma$ is larger than the cross section
$\sigma_{tz}(s)$ of the process $ep\rightarrow\gamma c\rightarrow
tz$. For $\varepsilon\leq0.08$ and $m_{\pi_{t}}\geq200GeV$,
$\sigma_{t\gamma}(s)$ and $\sigma_{tz}(s)$ at the $HERA$ collider
are smaller than 0.14$pb$ and 0.066$pb$, respectively. However, at
the $THERA$ collider with $\sqrt{s}=1000GeV$,
$\sigma_{t\gamma}(s)$ and $\sigma_{tz}(s)$ are in the ranges of
0.14$pb\sim1.37pb$ and $0.13pb\sim1.35pb$, respectively, for
$0.02\leq\varepsilon\leq0.08$ and $200GeV\leq
m_{\pi_{t}}\leq400GeV$.

If we assume that the $HERA$ collider with $\sqrt{s}=320GeV$ has a
yearly integrated luminosity of $ \pounds=160pb^{-1}$ and the
$THERA$ collider  with $\sqrt{s}=1000GeV$ has a yearly integrated
luminosity of $\pounds=470pb^{-1}$[6], then the yearly production
events of $t\gamma$ and $tz$ can be easily estimated. In most of
the parameter space of $TC2$ models, there may be only about 10 or
less of $t\gamma$ events or $tz$ events generated a year in the
$HERA$ collider, which is very difficult to detect. However, there
may be hundreds of $t\gamma$ events or $tz$ events to be generated
a year in the $THERA$ collider. For example, for
$m_{\pi_{t}}=300GeV$ and $\varepsilon=0.05$, the $THERA$ collider
can generate 252 $t\gamma$ events and 240 $tz$ events. Thus, the
effects of the neutral top-pion $\pi_{t}^{0}$ on $t\gamma$
production and $tz$ production may be detected at the $THERA$
collider.

\begin{figure}[htb]
\begin{center}
\epsfig{file=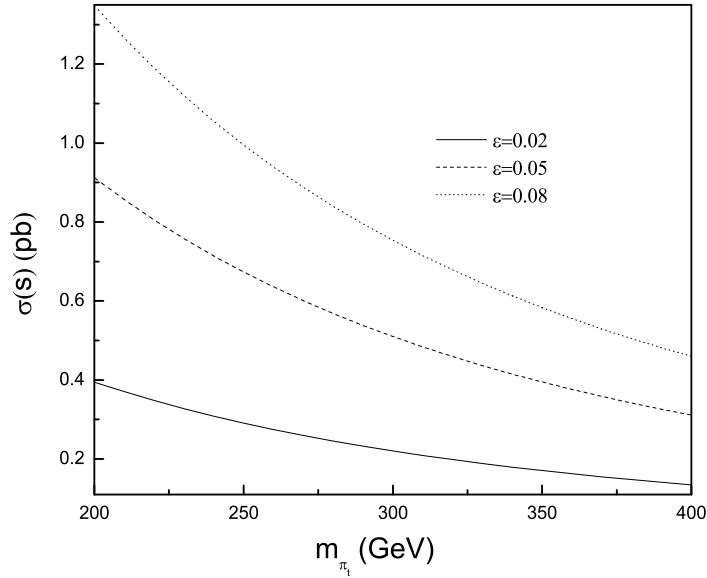,width=300pt,height=250pt} \vspace{-1.0cm}
\hspace{5mm} \caption{Same as Fig.4 but for the cross section
$\sigma_{tz}(s)$.} \label{ee}
\end{center}
\end{figure}

In conclusion, $TC2$ models predict the existence of the neutral
top-pion $\pi_{t}^{0}$, which can induce the anomalous top quark
couplings $tc\gamma$ and $tcz$ and further contribute to single
top quark production. In this letter, we calculated the
contributions of $\pi_{t}^{0}$ to $t\gamma$ production and $tz$
production via the processes $ep\rightarrow\gamma c\rightarrow
t\gamma$ and $ep\rightarrow\gamma c\rightarrow t z$ at the $HERA$
and $THERA$ colliders. We find that the cross sections of
$t\gamma$ and $tz$ production are very small at the $HERA$
collider. The effects of the neutral top-pion $\pi_{t}^{0}$ on
$t\gamma$ and $tz$ production can not be observed at the $HERA$
collider. However, $\pi_{t}^{0}$ exchange can generate several
hundred $t\gamma$ or $tz$ events at the $THERA$ collider.

\vspace{.5cm} \noindent{\bf Acknowledgments}

Dongqi Yu would like to thank Professor Xuelei Wang for useful
discussions. We thank the referee for carefully reading the
manuscript. This work was supported by the National Natural
Science Foundation of China (90203005) and the Natural Science
Foundation of the Liaoning Scientific Committee(20032101).

\end{document}